# Detection of Pulsed Gamma Rays Above 100 GeV from the Crab Pulsar

The VERITAS Collaboration

E. Aliu[1], T. Arlen[2], T. Aune[3], M. Beilicke[4], W. Benbow[5], A. Bouvier[3], S. M. Bradbury[6], J. H. Buckley[4], V. Bugaev[4], K. Byrum[7], A. Cannon[8], A. Cesarini[9], J. L. Christiansen[10], L. Ciupik[11], E. Collins-Hughes[8], M. P. Connolly[9], W. Cui[12], R. Dickherber[4], C. Duke[13], M. Errando[1], A. Falcone[14], J. P. Finley[12], G. Finnegan[15], L. Fortson[16], A. Furniss[3], N. Galante[5], D. Gall[12], K. Gibbs[5], G. H. Gillanders[9], S. Godambe[15], S. Griffin[17], J. Grube[11], R. Guenette[17], G. Gyuk[11], D. Hanna[17], J. Holder[18], H. Huan[19], G. Hughes[20], C. M. Hui[15], T. B. Humensky[19], A. Imran[21], P. Kaaret[22], N. Karlsson[16], M. Kertzman[23], D. Kieda[15], H. Krawczynski[4], F. Krennrich[21], M. J. Lang[9], M. Lyutikov[12], A. S Madhavan[21], G. Maier[20], P. Majumdar[2], S. McArthur[4], A. McCann[17,*], M. McCutcheon[17], P. Moriarty[24], R. Mukherjee[1], P. Nuñez[15], R. A. Ong[2], M. Orr[21], A. N. Otte[3,*], N. Park[19], J. S. Perkins[5], F. Pizlo[12], M. Pohl[20,25], H. Prokoph[20], J. Quinn[8], K. Ragan[17], L. C. Reyes[19], P. T. Reynolds[26], E. Roache[5], J. Rose[6], J. Ruppel[25], D. B. Saxon[18], M. Schroedter[5,*], G. H. Sembroski[12], G. D. Şentürk[1], A. W. Smith[7], D. Staszak[17], G. Tešić[17], M. Theiling[5], S. Thibadeau[4], K. Tsurusaki[22], J. Tyler[17], A. Varlotta[12], V. V. Vassiliev[2], S. Vincent[15], M. Vivier[18], S. P. Wakely[19], J. E. Ward[8], T. C. Weekes[5], A. Weinstein[2], T. Weisgarber[19], D. A. Williams[3], B. Zitzer[12]

[1] Department of Physics and Astronomy, Barnard College, Columbia University, NY 10027, USA

[2] Department of Physics and Astronomy, University of California, Los Angeles, CA 90095, USA

[3] Santa Cruz Institute for Particle Physics and Department of Physics, University of California, Santa Cruz, CA 95064, USA

[4] Department of Physics, Washington University, St. Louis, MO 63130, USA

[5] Fred Lawrence Whipple Observatory, Harvard-Smithsonian Center for Astrophysics, Amado, AZ 85645, USA

[6] School of Physics and Astronomy, University of Leeds, Leeds, LS2 9JT, UK

[7] Argonne National Laboratory, 9700 S. Cass Avenue, Argonne, IL 60439, USA

[8] School of Physics, University College Dublin, Belfield, Dublin 4, Ireland

[9] School of Physics, National University of Ireland Galway, University Road, Galway, Ireland

[10] Physics Department, California Polytechnic State University, San Luis Obispo, CA 94307, USA

[11] Astronomy Department, Adler Planetarium and Astronomy Museum, Chicago, IL 60605, USA

[12] Department of Physics, Purdue University, West Lafayette, IN 47907, USA

[13] Department of Physics, Grinnell College, Grinnell, IA 50112-1690, USA

[14] Department of Astronomy and Astrophysics, 525 Davey Lab, Pennsylvania State University, University Park, PA 16802, USA

[15] Department of Physics and Astronomy, University of Utah, Salt Lake City, UT 84112, USA

[16] School of Physics and Astronomy, University of Minnesota, Minneapolis, MN 55455, USA

[17] Physics Department, McGill University, Montreal, QC H3A 2T8, Canada

[18] Department of Physics and Astronomy and the Bartol Research Institute, University of Delaware, Newark, DE 19716, USA

[19] Enrico Fermi Institute, University of Chicago, Chicago, IL 60637, USA

[20] DESY, Platanenallee 6, 15738 Zeuthen, Germany

[21] Department of Physics and Astronomy, Iowa State University, Ames, IA 50011, USA

[22] Department of Physics and Astronomy, University of Iowa, Van Allen Hall, Iowa City, IA 52242, USA

[23] Department of Physics and Astronomy, DePauw University, Greencastle, IN 46135-0037, USA

[24] Department of Life and Physical Sciences, Galway-Mayo Institute of Technology, Dublin Road, Galway, Ireland

[25] Institut für Physik und Astronomie, Universität Potsdam, 14476 Potsdam-Golm,Germany

[26] Department of Applied Physics and Instrumentation, Cork Institute of Technology, Bishopstown, Cork, Ireland

*Authors to which correspondence should be addressed: nepomuk.otte@gmail.com (N.O.); mccann@hep.physics.mcgill.ca (A.M.); schroedter@veritas.sao.arizona.edu (M.S.)

We report the detection of pulsed gamma rays from the Crab pulsar at energies above 100 Gigaelectronvolts (GeV) with the VERITAS array of atmospheric Cherenkov telescopes. The detection cannot be explained on the basis of current pulsar models. The photon spectrum of pulsed emission between 100 Megaelectronvolts (MeV) and 400 GeV is described by a broken power law that is statistically preferred over a power law with an exponential cutoff. It is unlikely that the observation can be explained by invoking curvature radiation as the origin of the observed gamma rays above 100 GeV. Our findings require that these gamma rays be produced more than 10 stellar radii from the neutron star.

Pulsars were first discovered over 40 years ago (*1*), and are now believed to be rapidly rotating, magnetized neutron stars. Within the corotating magnetosphere, charged particles are accelerated to relativistic energies and emit non-thermal radiation from radio waves through gamma rays. While this picture reflects the broad scientific consensus, the details are still very much a mystery. For example, a number of models exist that can be distinguished from each other based on the location of the acceleration zone. Popular examples include the outer-gap model (*2-5*), the slot-gap model (*6, 7*), and the pair-starved polar-cap model (*8-10*). One way to better understand the dynamics within the magnetosphere is through observation of gamma rays emitted by the accelerated particles.

All of the detected gamma-ray pulsars in (*11*) exhibit a break in the spectrum between a few hundred MeV and a few GeV, with a rapidly fading flux above the break. The break energy is related to the maximum energy of the particles and to the efficiency of the pair production. Mapping the cutoff can help to constrain the geometry of the acceleration region, the gamma-ray radiation mechanisms, and the attenuation of gamma-rays. Previous measurements of the spectral break are statistically compatible with an exponential or sub-exponential cutoff, which is currently the most favored shape for the spectral break.

One of the most powerful pulsars in gamma rays is the Crab pulsar (*12, 13*), PSR J0534+220, which is the remnant of a historical supernova that was observed in 1054 A.D. It is located at a distance of 6500±1600 light years, has a rotation period of ~33 ms, a spin-down power of $4.6 \times 10^{38}$ erg $s^{-1}$ and a surface magnetic field of $3.8 \times 10^{12}$ G (*14*). Attempts to detect pulsed gamma rays above 100 GeV from the Crab pulsar began decades ago (*15*). Prior to the work reported here, the highest energy detection was at 25 GeV (*16*). At higher energies, near 60 GeV, only hints of pulsed emission have been reported in two independent observations (*16, 17*). Although measurements of the Crab pulsar spectrum are consistent, within the errors of the measurements, with a power law with an exponential cutoff at about 6 GeV (13), the flux measurements above 10 GeV are systematically higher than the fit with an exponential cutoff, hinting that the spectrum is indeed harder than a power law with exponential cutoff (*13, 16*). However, the sensitivity of the previous data was insufficient to allow a definite conclusion about the spectral shape.

We observed the Crab pulsar with VERITAS for 107 hours between September 2007 and March 2011. VERITAS is a ground-based gamma-ray observatory composed of an array of four atmospheric Cherenkov telescopes located in southern Arizona, USA (*18*). VERITAS has a trigger threshold of 100 GeV. Most of the data, 77.7 hours, were recorded after the relocation in summer 2009 of one of the VERITAS telescopes, which resulted in a lower energy threshold and better sensitivity of the array. We processed the recorded atmospheric shower images with a standard moment analysis (*19*) and calculated the energy and arrival direction of the primary particles (*20*). We then rejected events caused by charged cosmic-ray events. For gamma rays, the distribution of the remaining, or selected, events as a function of energy peaks at 120 GeV. In the pulsar analysis, for each selected event, we first transformed the arrival time to the barycenter of the solar system and then calculated the spin phase of the Crab pulsar from the barycentered time using contemporaneously measured spin-down parameters (*21*). All steps in the analysis have been cross-checked by an independent software package and are explained in detail in the appendix. We applied the H-Test (*22*) to test for periodic emission at the frequency of the Crab pulsar (Appendix). This yielded a test value of 50, corresponding to a significance of 6.0 standard deviations that pulsed emission is present in the data.

The phase-folded event distribution, hereafter pulse profile, of the selected VERITAS events is shown in Figure 1. The most significant structures are two pulses with peak amplitudes at phase 0.0 and phase 0.4. These coincide with the locations of the main pulse and interpulse, hereafter P1 and P2, which are the two main features in the pulse profile of the Crab pulsar throughout the electromagnetic spectrum. We characterized the pulse profile using an unbinned maximum-likelihood fit (Appendix). In the fit, the pulses were modeled with Gaussian functions, and the background was determined from the events that fell between phases 0.43 and 0.94 in the pulse profile (referred to as the off-pulse region). The positions of P1 and P2 in the VERITAS data thus lie at the phase values $-0.0026 \pm 0.0028$ and $0.3978 \pm 0.0020$, respectively and are shown by the vertical lines (Fig. 1). The full widths at half maximum (FWHM) of the fitted pulses are $0.0122 \pm 0.0035$ and $0.0267 \pm 0.0052$, respectively. The pulses are narrower by a factor two to three than those measured by Fermi-LAT - at 100 MeV – (*13*) (Fig. 1).

If gamma rays observed at the same phase are emitted by particles that propagate along the same magnetic field line (*23*) and the electric field in the acceleration region is homogeneous, then a possible explanation of the observed narrowing is that the region where acceleration occurs tapers towards the neutron star. However, detailed calculations are necessary to explain fully the observed pulse profile.

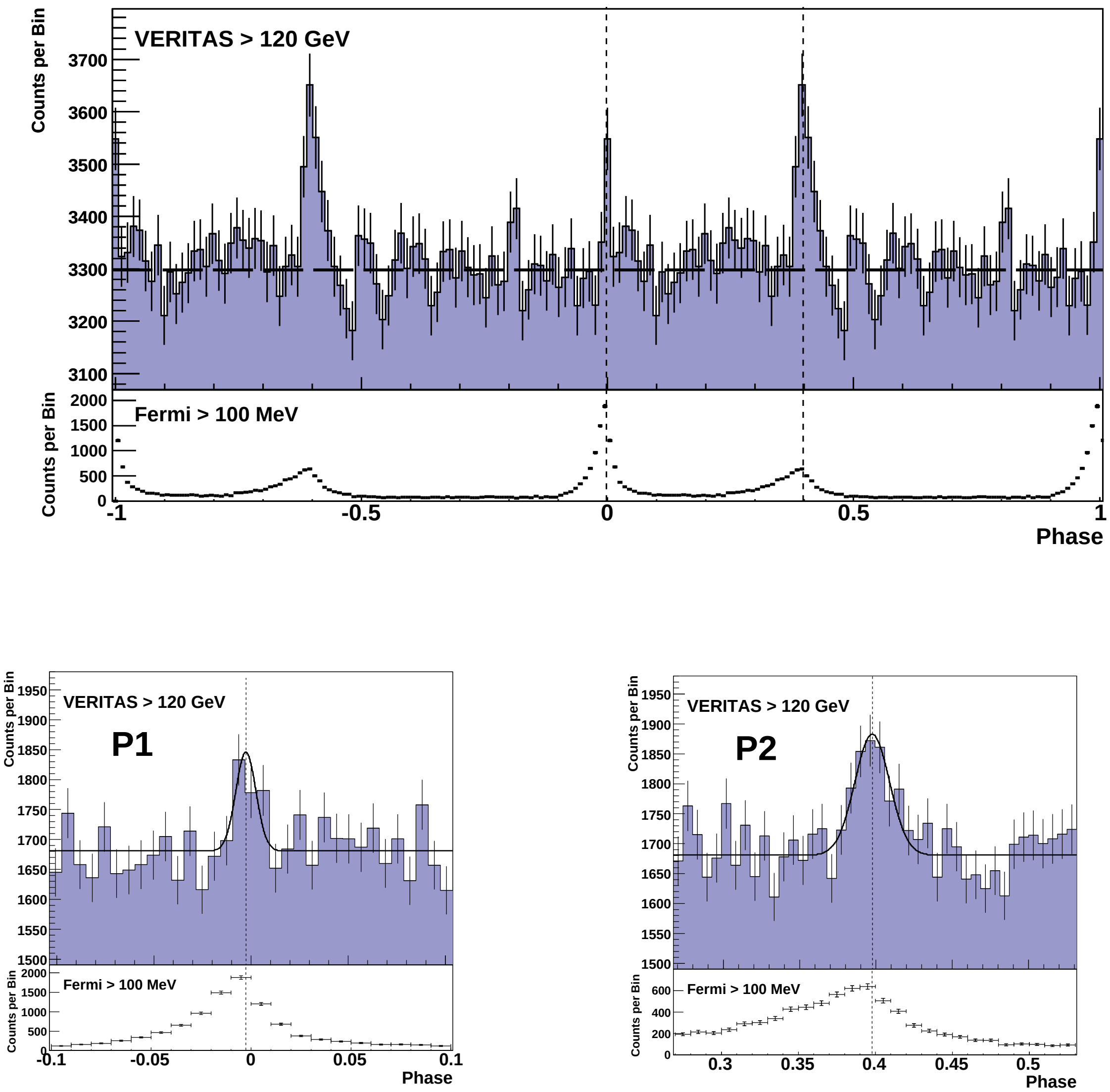

**Fig. 1.** Pulse profile of the Crab pulsar. Phase 0 is the position of P1 in radio. The shaded histograms show the VERITAS data. The pulse profile in the upper panel is shown twice for clarity. The dashed horizontal line in the upper panel shows the background level estimated from data in the phase region between 0.43 and 0.94. The lower panels show expanded views of the pulse profile with a finer binning than in the upper panel and are centered at P1 and P2, which are the two dominant features in the pulse profile of the Crab pulsar. The data above 100 MeV from the Fermi-LAT (*13*) are shown beneath the VERITAS profile. The vertical dashed lines in the panels mark the best-fit peak positions of P1 and P2 in the VERITAS data. The solid black line in the lower panels shows the result of an unbinned maximum-likelihood fit of Gaussian functions to the VERITAS pulse profile (described in text). The peak positions between the Fermi-LAT and the VERITAS data agree within uncertainties.

Along with the observed differences in the pulse width, the amplitude of P2 is larger than P1 in the profile measured with VERITAS, in contrast to what is observed at lower gamma-ray energies where P1 dominates (Fig. 1). It is known that the ratio of the pulse amplitudes changes as a function of energy above 1 GeV (*13*) and becomes near unity for the pulse profile integrated above 25 GeV (*16*). In order to quantify the relative intensity of the two peaks above 120 GeV, we integrated the pulsed excess between phase -0.013 and 0.009 for P1 and between 0.375 and 0.421 for P2. This is the ±2 standard deviation interval of each pulse as determined from the maximum-likelihood fit. The ratio of the excess events and thus the intensity ratio of P2/P1 is 2.4 ± 0.6. If one assumes that the differential energy spectra of P1 and P2 above 25 GeV can each be described with a power law, $F(E) \sim E^{\alpha}$, and that the intensity ratio is exactly unity at 25 GeV (*16*), then the spectral index α of P1 must be smaller than the spectral index of P2 by $\alpha_{P2} - \alpha_{P1} = 0.56 \pm 0.16$.

We measured the gamma-ray spectrum above 100 GeV by combining the pulsed excess in the phase regions around P1 and P2. This can be considered a good approximation of the phase-averaged spectrum because no "bridge emission", which is observed at lower energies, is seen between P1 and P2 in the VERITAS data. However, the existence of a constant flux component that originates in the magnetosphere cannot be excluded and would be indistinguishable from the gamma-ray flux from the nebula. Figure 2 shows the VERITAS phase-averaged spectrum together with measurements made with Fermi-LAT and MAGIC. In the energy range between 100 GeV and 400 GeV measured by VERITAS, the energy spectrum is well described by a power law $F(E) = A(E/150\ \mathrm{GeV})^{\alpha}$, with $A = (4.2 \pm 0.6_{stat} +2.4_{syst} -1.4_{syst}) \times 10^{-11}\ \mathrm{TeV^{-1}\ cm^{-2}\ s^{-1}}$ and $\alpha = -3.8 \pm 0.5_{stat} \pm 0.2_{syst}$. At 150 GeV, the flux from the pulsar is approximately 1 % of the flux from the nebula. The detection of pulsed gamma-ray emission between 200 GeV and 400 GeV, the highest energy flux point, is only possible if the emission region is at least 10 stellar radii from the star's surface (*24*).

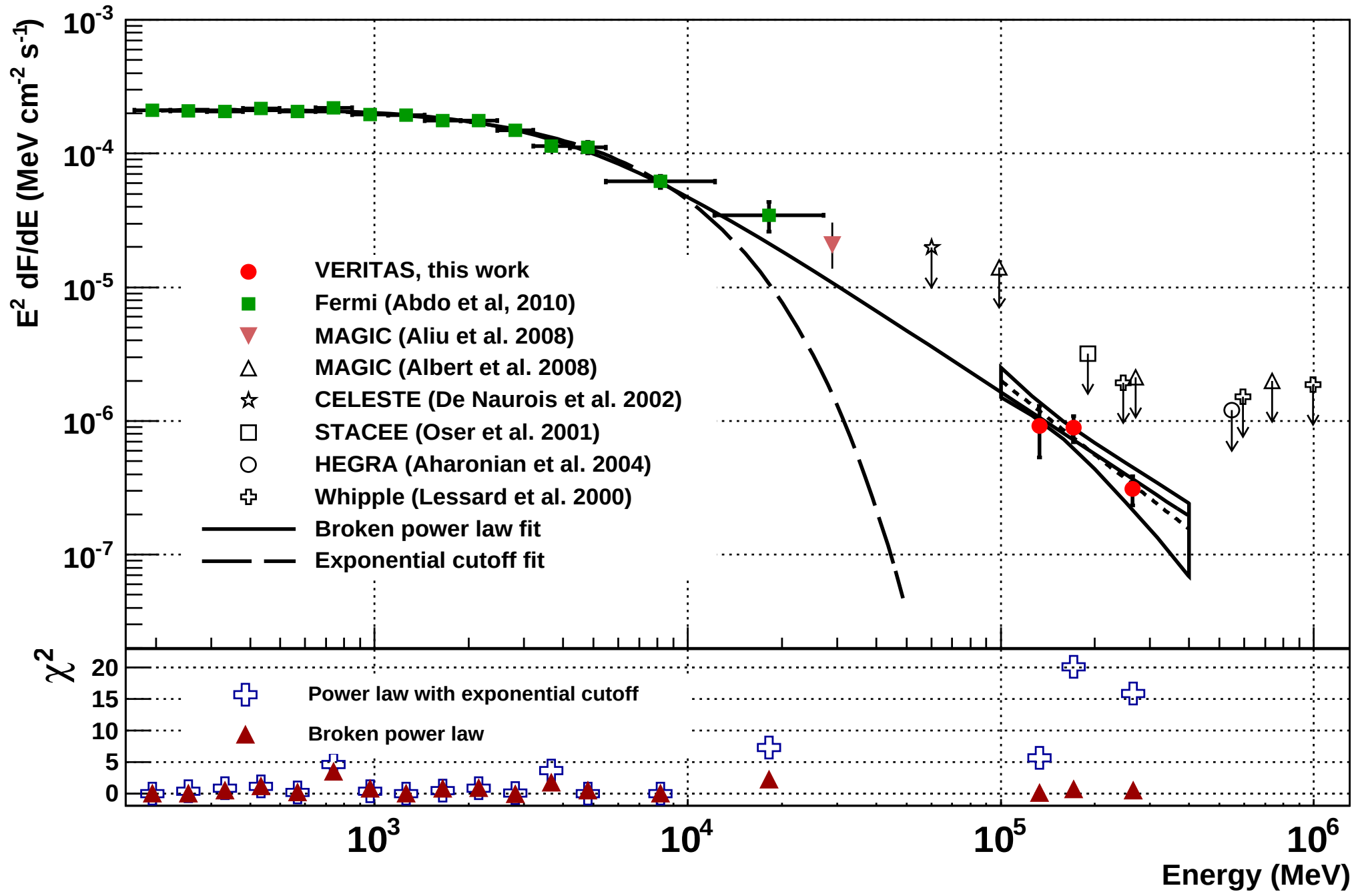

**Fig. 2.** Spectral energy distribution (SED) of the Crab pulsar in gamma rays. VERITAS flux measurements are shown by the solid red circles, Fermi-LAT data (*13*) by green squares, and the MAGIC flux point (*16*) by the solid triangle. The empty symbols are upper limits from CELESTE (*25*), HEGRA (*26*), MAGIC (*17*), STACEE (*27*), and Whipple (*29*). The bowtie and the enclosed dotted line give the statistical uncertainties and the best-fit power-law spectrum for the VERITAS data using a forward-folding method. The result of a fit of the VERITAS and Fermi-LAT data with a broken power law is given by the solid line and the result of a fit with a power-law spectrum multiplied with an exponential cutoff is given by the dashed line. Below the SED we plot $\chi^2$ values to visualize the deviations of the best-fit parametrization from the Fermi-LAT and VERITAS flux measurements.

Combining the VERITAS data with the Fermi-LAT data we can place a stringent constraint on the shape of the spectral turnover. The previously favored spectral shape of the Crab pulsar above 1 GeV was an exponential cutoff $F(E) = A(E/E_0)^{\alpha}\exp(-E/E_c)$, which is a good parametrization of the Fermi-LAT (*13*) and MAGIC (*16*) data. The Fermi-LAT and MAGIC data can be equally well parametrized by a broken power law but those data are not sufficient to distinguish significantly between a broken power law and an exponential cutoff. The VERITAS data, on the other hand, clearly favor a broken power law as a parametrization of the spectral shape. The fit of the VERITAS and Fermi-LAT data with a broken power law of the form $A(E/E_0)^{\alpha}/[1 + (E/E_0)^{\alpha-\beta}]$ results in a $\chi^2$ value of 13.5 for 15 degrees of freedom with the fit parameters $A = (1.45 \pm 0.15_{stat}) \times 10^{-5}$ TeV$^{-1}$ cm$^{-2}$ s$^{-1}$, $E_0 = 4.0 \pm 0.5_{stat}$ GeV, $\alpha = -1.96 \pm 0.02_{stat}$ and $\beta = -3.52 \pm 0.04_{stat}$ (Fig. 2). A corresponding fit with a power law and an exponential cutoff yields a $\chi^2$ value of 66.8 for 16 degrees of freedom. The fit probability of $3.6 \times 10^{-8}$ derived from the $\chi^2$ value excludes the exponential cutoff as a viable parametrization of the Crab pulsar spectrum.

The detection of gamma-ray emission above 100 GeV provides strong constraints on the gamma-ray radiation mechanisms and location of the acceleration regions. Assuming a balance between acceleration gains and radiative losses by curvature radiation, the break in the gamma-ray spectrum is expected to be at $E_{br}$ = 150 GeV $\eta^{3/4}$ sqrt($\xi$), where $\eta$ is the acceleration efficiency ($\eta < 1$) and $\xi$ is the radius of curvature in units of the light-cylinder radius (28, Appendix). Only in the extreme case of an acceleration field that is close to the maximum allowed value and a radius of curvature that is close to the light-cylinder radius would it be possible to produce gamma-ray emission above 100 GeV with curvature radiation. It is, therefore, unlikely that curvature radiation is the dominant production mechanism of the observed gamma-ray emission above 100 GeV. A plausible different radiation mechanism is inverse-Compton scattering that has motivated previous searches for pulsed VHE emission, e.g. (29). With regard to the overall gamma-ray production, two possible interpretations are that either one emission mechanism different from curvature radiation dominates at all gamma-ray energies or that a second mechanism, becomes dominant above the spectral break energy. It might be possible to distinguish between the two scenarios with higher-resolution spectral measurements above 10 GeV.

Acknowledgments: This research is supported by grants from the U.S. Department of Energy, the U.S. National Science Foundation and the Smithsonian Institution, by NSERC in Canada, by Science Foundation Ireland (SFI 10/RFP/AST2748) and by STFC in the U.K. We acknowledge the excellent work of the technical support staff at the Fred Lawrence Whipple Observatory and at the collaborating institutions in the construction and operation of the instrument. N. O. was supported in part by a Feodor-Lynen fellowship of the Alexander von Humboldt Foundation. We are grateful to M. Roberts and A. Lyne for providing us with Crab-pulsar ephemerides before the public ones became available.

# Appendix

In this appendix to the Crab pulsar detection above 100 GeV with VERITAS, we give details about the instrument, methods, and results. The document is structured in three sections. In section 1, we describe the VERITAS array of atmospheric Cherenkov telescopes, detail the data set, and explain the data reduction and event reconstruction. In section 2, we explain the pulsar analysis and the spectral reconstruction. In section 3, we discuss the analysis of the combined VERITAS and Fermi data sets that results in the exclusion of an exponential cutoff of the energy spectrum. Lastly, in section 4, we explain why it is unlikely that curvature radiation is the emission mechanism responsible for the pulsed gamma-ray emission above 100 GeV.

## 1. Observations of the Crab Pulsar with VERITAS and Event Reconstruction

VERITAS, the Very Energetic Radiation Imaging Telescope Array System, is an array of four imaging atmospheric Cherenkov telescopes located in southern Arizona, USA. Each of the four telescopes has a Davies-Cotton arrangement of 350 identical, hexagonal mirror facets yielding a 12 m diameter collector with f/D=1. Located in the focal plane of each telescope is a pixelated camera consisting of 499 photomultiplier tubes (PMTs), each with an angular size of 0.15 degrees. The camera images particle showers in the atmosphere by measuring a Cherenkov-light flash of a few nanoseconds duration. The recorded air showers are initiated by gamma rays or charged cosmic rays. For an event to trigger the readout, a coincident signal has to be detected in at least two telescopes. At the telescope level, the trigger requirement is a signal of more than 6 photoelectrons in three or more neighboring pixels within a coincidence window of 9 nanoseconds. If two telescopes trigger within 50 nanoseconds, the readout is triggered, in which case the PMT signals of all cameras that are digitized with 500 MSample/s flash analog to digital converters (FADCs) are written to hard disk. While the trigger rate at the single-telescope level is typically several kilohertz and dominated by triggers caused by fluctuations in the night sky background, the additional coincidence requirement of a two-fold telescope coincidence reduces the accidental rate to less than 10 Hz and yields a cosmic ray trigger rate of about 230 Hz. For a more detailed description of VERITAS we refer the interested reader to (18).

The Crab is a regular observation target of VERITAS, mainly for the purpose of monitoring the array performance by means of the strong gamma-ray emission of the Crab Nebula. After evidence for pulsed emission was seen in 45 hours of data that were recorded between 2007 and 2010, a deep 62-hour observation was carried out on the Crab pulsar between September 2010 and March 2011. The observations were made in “wobble” mode in which the source is offset from the center of the field of view of the cameras by 0.5 degree. This is the standard observing mode for point sources with VERITAS and allows simultaneous background measurements. For the spectral analysis, the background is estimated from events that fall in a region of the pulse profile where no pulsed emission is expected from the pulsar. After eliminating data taken under variable or poor sky conditions or affected by technical problems, the total analyzed data set comprises 107 hours of observations (97 hours dead-time corrected) carried out with all four telescopes. In order to obtain a trigger threshold of 100 GeV, the observations were restricted to zenith angles < 25 degrees.

Data Processing and Event Selection

The analysis of an event is performed by first processing the images recorded in each telescope and then combining the images to reconstruct the characteristics of the primary particle. In the first step, the PMT signals in each camera are corrected for gain differences between the PMTs, and pixels containing only noise are removed. If after this cleaning procedure an image remains with a total of 20 or more photoelectrons, it is parametrized with a principal-moment analysis called a Hillas analysis (19). If an image is found in at least two telescopes, the parametrized images in the triggered telescopes are then combined and the event is characterized by calculating the position of the maximum Cherenkov emission, the projected impact point on the ground, and the direction and energy of the primary particle.

After event reconstruction, a selection is performed to reject events caused by charged cosmic rays. The selection criteria were optimized a priori for highest sensitivity by assuming a simple power-law energy spectrum, $F(E) \sim E^{\alpha}$, for the Crab pulsar with an index $\alpha = -4$ and a flux normalization at 100 GeV that is equivalent to the extrapolation of the broken power-law fit of the Fermi-LAT data below 10 GeV. The optimization of the selection criteria took into account the fact that the gamma-ray signal from the pulsar is contaminated not only by charged cosmic ray events but also by gamma rays coming from the Crab Nebula. In fact, after event selection, about half of the remaining background events are due to gamma rays from the Crab Nebula. The selection parameters and values (30) are the angular separation between the source location and the shower direction, i.e., theta (<0.27 degree), mean scaled width (<1.17), mean scaled length (<1.35), and height of shower maximum (>6.6 km). The results presented were confirmed by a separate analysis of the data made using an independent analysis package.

## 2. Pulsar and Spectral Analysis of the VERITAS Data

Pulsar analysis

For the pulsar analysis, the arrival times of the selected events are transformed to the barycenter of the solar system. The barycentering was done with two custom codes and the tempo2 pulsar timing package (31). The agreement in the barycentered times between all the codes is better than 10 microseconds. The event times themselves are derived from four independent GPS clocks and have an accuracy that is better than 1 microsecond. After barycentering, the phase of the Crab pulsar is calculated for each event using contemporaneous ephemerides of the Crab pulsar that are published monthly by the Jodrell Bank telescope (21). In these ephemerides, phase zero is aligned with the position of the peak of P1 in radio at 608 MHz. The distribution of the calculated phases is shown in Fig. S1. A clear excess is evident at the position of P1 (phase 0.0) and the position of P2 (phase 0.4). These are the same locations where pulsed emission is observed in radio, optical, X-rays, and gamma rays. In order to assess the significance of the pulsed emission, we use the H-Test (22) that does not make an a priori assumption about the shape of the pulse profile and is applied to the unbinned data. The test result is 50, which translates into a statistical significance of 6.0 standard deviations that pulsed emission is present in the data.

Figure S2 shows how the number of pulsed excess events and the statistical significance grow versus the total number of events used in the analysis. The phase regions chosen for these figures are the same regions used in the spectrum reconstruction. The excess number of events grows linearly and the significance grows following a square-root behavior, as expected for a constant gamma-ray source. The pulsed excess is 1211 ± 138 events, and the total number of events after selection cuts is 268,949 events, of which about 50% are gamma rays from the Crab Nebula.

The observed excess cannot be explained by triggers caused by the optical emission from the Crab pulsar. During one rotation of the pulsar, each VERITAS telescope detects about 104 photoelectrons from P1 and P2. Thus the average rate is $3x10^{-3}$ photoelectrons per nanosecond for a pulse duration of 3 ms. If the optical emission is to trigger one pixel in the camera, it would require about 6 photoelectrons to pile up within 3 nanoseconds. The probability that this happens during one rotation of the pulsar is about $10^{-10}$. During the 107 hours of VERITAS observation, any given pixel triggered with a probability of less than $10^{-5}$. It can be safely concluded that none of the about 1100 pulsed excess events was triggered by the optical emission from the Crab pulsar. This estimate does not include the array trigger and three-nearest-neighbor requirement of the telescope trigger, both of which reduce the probability of triggers due to the optical emission of the Crab pulsar even further.

Pulse Profile

Figure S1 shows the pulse profile of all events surviving the event reconstruction and selection criteria. The analysis threshold is 120 GeV, defined as the peak in the differential trigger rate for a simple power-law spectrum with index $\alpha = -3.8$, which is the best-fit spectrum to the VERITAS data. Comparing this pulse profile with the pulse profile at 100 MeV it is evident that the pulses at 120 GeV are much narrower than at lower energies.

In order to quantify the peak positions and the widths of the two peaks, we performed an unbinned maximum-likelihood fit in which the two peaks are described by Gaussian functions. As probability density function PDF we used

$$PDF(\varphi) = A_1 / \mathrm{sqrt}(2*\pi*\sigma_1^2) * \exp[-(\varphi-\mu_1)^2/(2\,\sigma_1^2)]$$
$$+ A_2 / \mathrm{sqrt}(2*\pi*\sigma_2^2) * \exp[-(\varphi-\mu_2)^2/(2\,\sigma_2^2)] + B,$$

which is normalized with $1/(A_1 + A_2 + B)$. $A_x$ and $\sigma_x$ are the number of events in the pulse and the standard deviation of the pulse, respectively. B is the number of background events. We define as likelihood function $L = -2 \log( \Pi_i [PDF(\varphi_i)] )$, which is minimized to find the best fit values for $A_x$ and $\sigma_x$. In the fit, the background is fixed to the average number of counts between phase 0.43 and 0.94, while the other parameters are kept free.

The result is a best-fit peak position of $-0.0026 \pm 0.0028$ for P1 and $0.3978 \pm 0.0020$ for P2. The full widths at half maximum (FWHM) of the pulses are $0.0122 \pm 0.0035$ and $0.0267 \pm 0.0052$ for P1 and P2, respectively. The fitted pulse profile is shown as a solid black line on top of the binned pulse profile in the lower panels of Figure 1 in the main paper. The uncertainties have been determined by simulating pulse profiles in which the best-fit parameters are used as a template. The total number of events for each simulated pulse profile is drawn from a Poisson distribution with a mean equal to the number of events in the VERITAS pulse profile. Each simulated pulse profile is fitted with the likelihood method and the best-fit parameters stored. The uncertainty in one parameter is then given by the root mean square of the distribution of the simulated best-fit values for that parameter.

While the present data are in good agreement with symmetric Gaussian functions, this does not exclude the possibility that, with a more sensitive data set, asymmetry of the pulses may be found in the future.

Besides the excess at P1 and P2, the excess with the next highest significance is at phase 0.81. The pre-trial statistical significance of the excess is 2.5 standard deviations, which is consistent with expected random fluctuations for the given binning of the pulse profile.

Energy Spectrum

The energy spectrum of the Crab pulsar above 100 GeV has been determined in two different ways. In both cases, the flux is derived by integrating the pulsed excess between phase -0.013 and 0.009 for P1 and between 0.375 and 0.421 for P2. The significance of the pulsed excess in these regions is 4.7 standard deviations for P1 and 7.9 standard deviations for P2. The background is estimated from the events with phases between 0.43 and 0.94, which includes both cosmic-ray background and gamma rays from the Crab Nebula. In the first method, a true energy spectrum is assumed and then folded through the instrumental response of VERITAS (forward folding). The result is then compared with the measured pulsed excess counts distribution and the $\chi^2$ deviation is calculated. The parameters of the true energy spectrum are iteratively varied to provide a $\chi^2$ profile which yields best-fit parameters and parameter uncertainties. The best-fit spectrum is shown in Figure S3 by the dotted line, $(4.2 \pm 0.6_{stat} +2.4_{syst} -1.4_{syst})$ x $10^{-11}$ $TeV^{-1}$ $cm^{-2}$ $s^{-1}$ $(E/150\ GeV)^{-3.8 \pm 0.5(stat) \pm 0.2(syst)}$. The bowtie gives the statistical uncertainties of the forward-folding method.

In the second method, the photon flux is calculated in bins of reconstructed energy. The energy bias is corrected for by re-weighting the effective area with a different spectral index and iterating the index until it converges (32). The flux points derived in this way are shown in Figure S3, as well as the fit to the data points of a simple power law, $F(E) = A(E/200\ GeV)^{\alpha}$, given by the solid black line, $(1.4 \pm 0.2_{stat} +0.8_{syst} -0.5_{syst})$ x $10^{-11}$ $TeV^{-1}$ $cm^{-2}$ $s^{-1}$ $(E/200\ GeV)^{-3.8 \pm 0.5(stat) \pm 0.2(syst)}$, which is in good agreement with the first spectral reconstruction method. In each of the spectral reconstruction methods, the flux normalization is given at the energy where the correlations between the flux normalization and the spectral index are minimal. Figure S4 shows some more details of the forward-folding method. Hereafter, we will only discuss the results where the VERITAS data have been forward folded.

Systematic uncertainties affecting the spectral measurement are dominated by uncertainties in the Cherenkov light production, changes in the transmission in the atmosphere, and uncertainties in the optical throughput of the telescopes. The impact of the systematic uncertainties in the energy scale on the spectral reconstruction was estimated with Monte Carlo simulations in which the optical efficiency of VERITAS was changed by ±15%. The main conclusions of the paper are not affected by the systematic uncertainty.

A fit with a simple power law is a good description of the VERITAS data. We cannot, however, exclude the possibility that the energy spectrum above 100 GeV is in reality a narrow peak whose nature is distinct from the emission observed at lower energies.

## 3. Combined Fit of the Fermi-LAT and VERITAS Data

For the combined fit of the Fermi-LAT and VERITAS data, we fit the phase-averaged Crab pulsar flux points from the Fermi-LAT as found in (13) while simultaneously doing a forward-folded fit of the VERITAS distribution of pulsed excess events. The data have been fit with two functions, a simple power law with an exponential cutoff and a broken power law. In the case of the power law with exponential cutoff, $F(E) = A(E/6\ GeV)^{\alpha}\exp(-E/E_c)$, the best-fit parametrization is $A = (7.3 \pm 0.5)$ $x10^{-6}$

$\mathrm{TeV}^{-1}\ \mathrm{cm}^{-2}\ \mathrm{s}^{-1}$, $\alpha = -1.95 \pm 0.02_{stat}$ and $E_c = (5.5 \pm 0.6_{stat})$ GeV. The fit is shown by the dashed line in Figure 2 in the main paper. The $\chi^2$ value is 66.8, which for 16 degrees of freedom yields a fit probability of $3.6 \times 10^{-8}$. The high $\chi^2$ value is dominated by the VERITAS data and rules out the possibility that an exponential cutoff describes the energy spectrum of the Crab pulsar above 10 GeV. We note that in case of an exponential cutoff, the curvature in the Fermi-LAT data constrains the expected flux above 100 GeV to be consistent with zero for all practical purposes. Therefore, the detection of gamma-ray emission above 100 GeV already rules out the exponential cutoff with the same statistical significance level at which the signal is detected.

As an alternative, the Fermi and VERITAS data have been fit with a broken power law, $F(E) = A\ (E/E0)^{\alpha} / [1 + (E/E_0)^{\alpha-\beta}]$. Based on the $\chi^2$ value of 13.5 for 15 degrees of freedom, it can be concluded that a broken power law is a good description of the combined Fermi-LAT and VERITAS data. The best-fit broken power law is shown by the solid line in Figure 2 in the main paper. The parametrization is $A =(1.45 \pm 0.15_{stat}) \times 10^{-5}\ \mathrm{TeV}^{-1}\mathrm{cm}^{-2}\mathrm{s}^{-1}$, $E_0=(4.0 \pm 0.5_{stat})$ GeV, $\alpha = -1.96 \pm 0.02_{stat}$ and $\beta = -3.52 \pm 0.04_{stat}$.

A fit of the Fermi-LAT and VERITAS data with a log-parabola function also results in a good parametrization of the data above the spectral break but fails to describe adequately the hard spectrum below 500 MeV.

**4. Curvature Radiation above 100 GeV**

Here we estimate the maximum spectral-break energy of gamma-rays that can be emitted by curvature radiation in the magnetosphere of the Crab pulsar by following (28). The curvature-radiation spectrum emitted by electrons with a Lorentz factor $\gamma$ has a break at energy $E_{br} = 3/2\ \hbar\ \gamma^3\ c/R_C$, where $R_C$ is the radius of curvature of the magnetic field lines along which the electrons propagate. The maximal Lorentz factor $\gamma$ can be estimated by assuming that the accelerating electric field is a fraction $\eta$ of the magnetic field B and that the acceleration is balanced by radiative losses:

$$e\ c\ \eta\ B = 2/3\ e^2\ \gamma^4\ c\ (1/R_C)^2 \rightarrow \gamma = (3/2\ (\eta\ B)/e\ R_C^2)^{1/4},$$

where e is the elementary charge of the electron and c is the speed of light.
It then follows that

$$E_{br} = 3/2\ (\hbar\ c)/R_C\ \gamma^3 = (3/2)^{7/4}\ \hbar\ c\ (R_C)^{1/2}\ (\eta\ B/e)^{3/4} \qquad (1)$$

$R_C$ can be expressed in units of the radius of the light cylinder $R_L$, $R_C = \xi\ R_L = \xi\ c\ P / 2\ \pi$, where, P = 33 ms is the period of the Crab pulsar, and $\xi$ is a dimensionless scaling parameter. If, furthermore, B is replaced by the radial distribution of the magnetic field of a dipole $B = B_{NS}(R_{NS}/R)^3$, where $B_{NS} = 3.7 \times 10^{12}$ Gauss is the surface magnetic field of the neutron star and $R_{NS}$ = 10 km is the radius of the star, then it follows that $E_{br}$ is

$$\begin{aligned} E_{\mathrm{br}} &= (3\pi)^{7/4}\frac{\hbar}{(ce)^{3/4}}\eta^{3/4}\sqrt{\xi}\,\frac{B_{\mathrm{NS}}^{3/4}R_{\mathrm{NS}}^{9/4}}{P^{7/4}} \\ &= 150\,\mathrm{GeV}\,\eta^{3/4}\sqrt{\xi} \end{aligned} \qquad (2)$$

If the acceleration is limited by radiation reaction, the emission observed by VERITAS above 100 GeV can only be explained by curvature radiation if the break energy is also at or above 100 GeV. Such a high break energy is achieved only in the extreme case of an accelerating electric field that is close to the maximum allowed value, i.e. $\eta \approx 1$, and a radius of curvature that is on the order of the light cylinder radius, i.e.

ξ ~ 1. According to our present understanding of pulsar magnetospheres it is unlikely to find such an extreme combination within the light cylinder and, therefore, a different emission mechanism has to be invoked, for example, inverse-Compton scattering.

**Fig. S1**

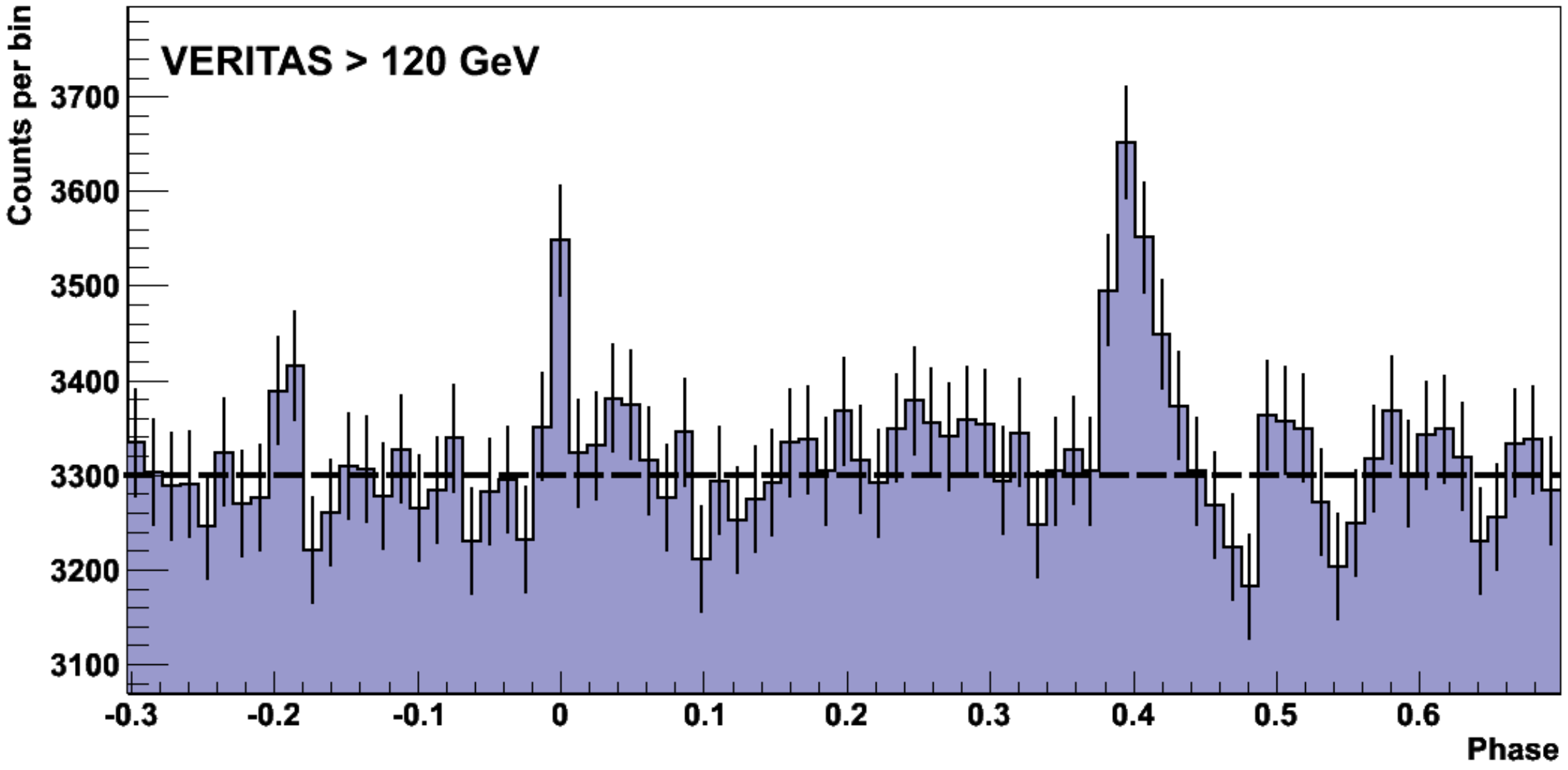

Pulse profile of the VERITAS events after applying selection criteria. See text and main paper for further discussions.

**Fig. S2**

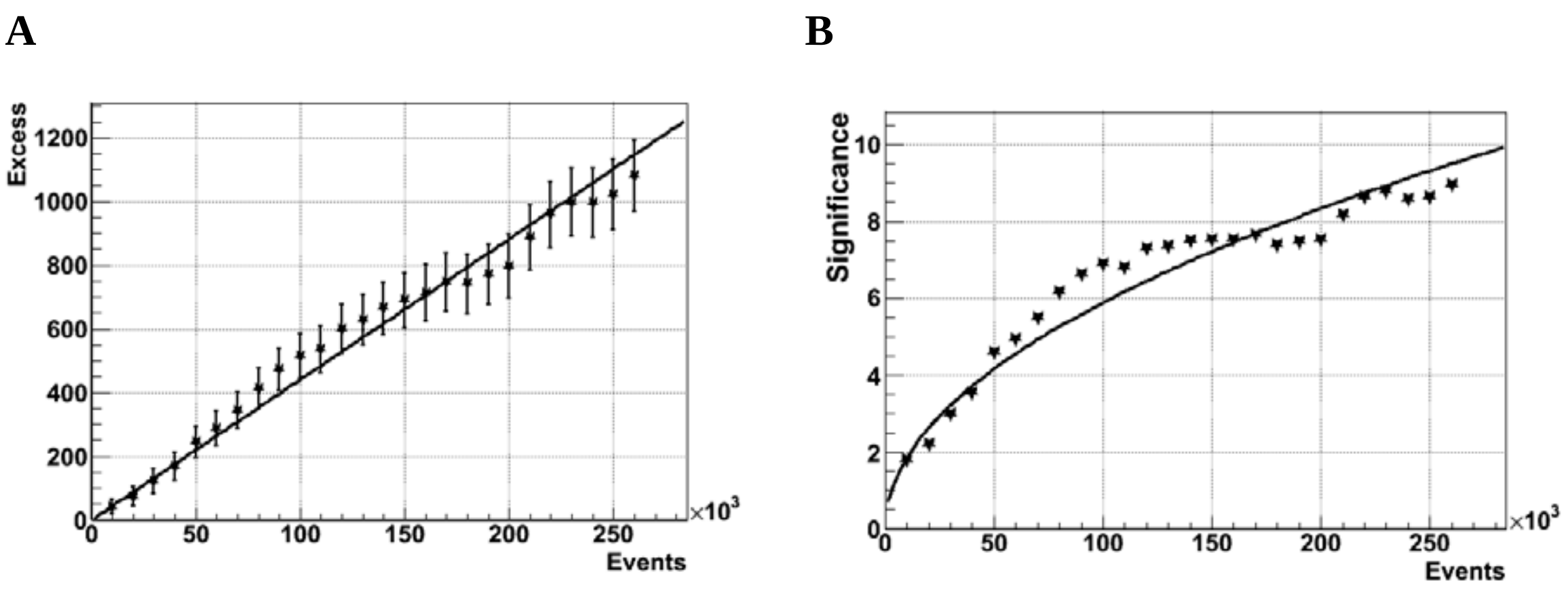

**A** Pulsed excess number as a function of the total number of accumulated events surviving the event reconstruction and selection criteria. The solid line gives the best fit with a linear function. **B** Growth of significance. The solid line gives the best fit with a square-root function. In both cases, the observed behavior is that of a constant gamma-ray source, i.e. linear growth of excess events and square-root growth of significance. Note that the data span four observing seasons (years) with greatly varying exposure per season. These figures show how the excess and significance grow in the phase regions that are used to compute the energy spectrum.

**Fig. S3**

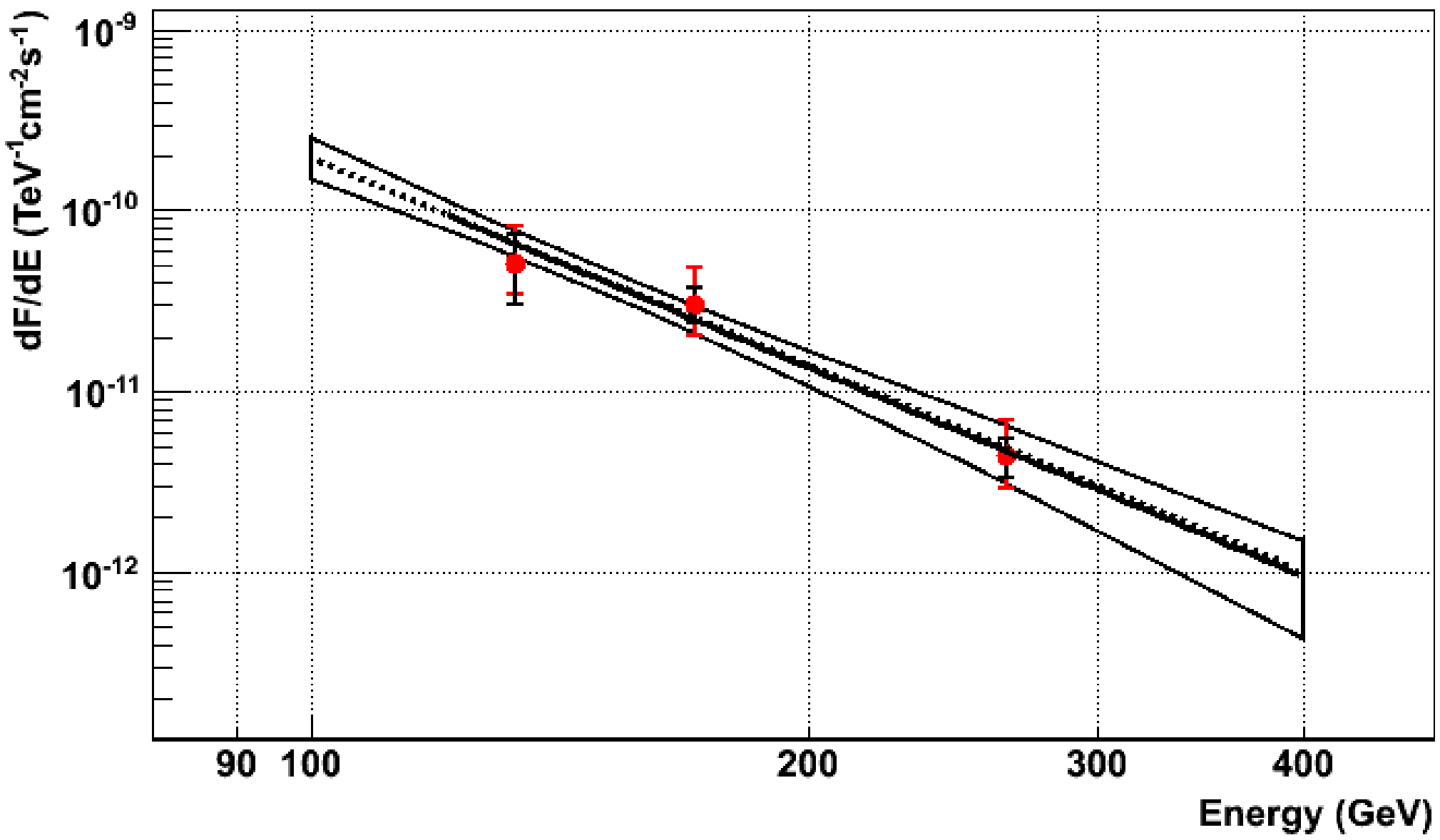

Differential energy spectrum of the Crab pulsar above 100 GeV. The black error bars on the VERITAS data show statistical uncertainty while the red error bars show systematic uncertainty. The bowtie and the enclosed dotted line give the statistical uncertainties and the best-fit power-law spectrum for the VERITAS data using a forward-folding method. The solid line gives the fit result of the three data points with a simple power law. See text for details.

**Fig. S4**

**A**

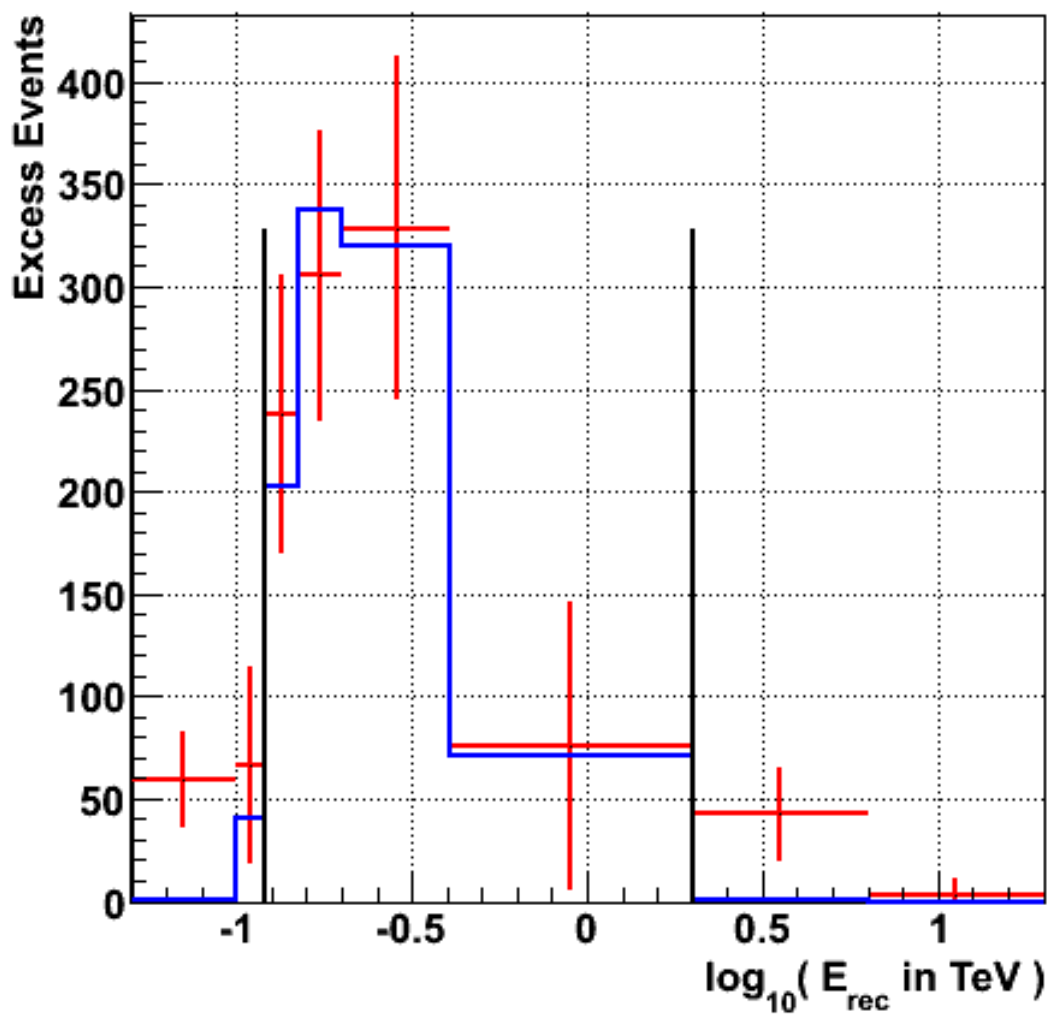

**B**

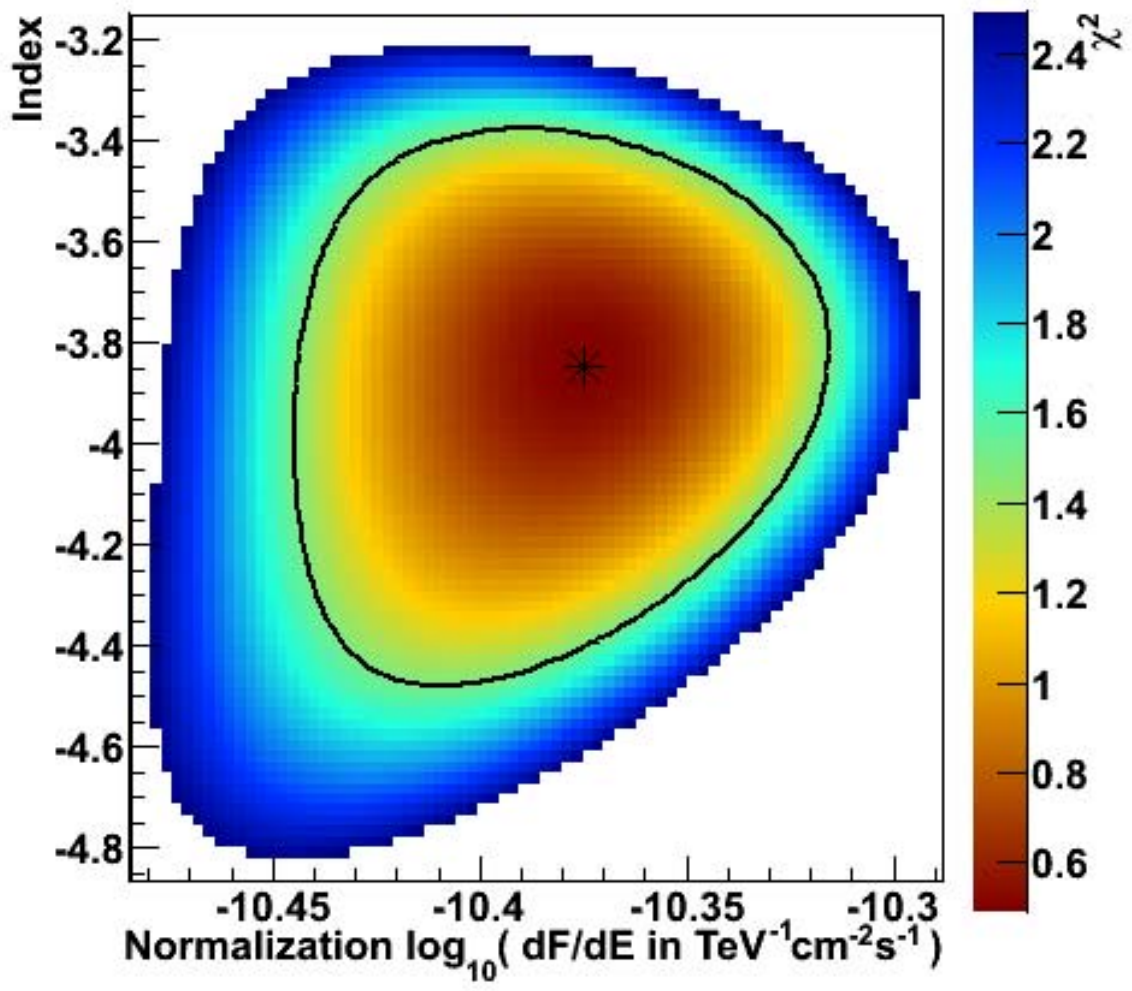

**C**

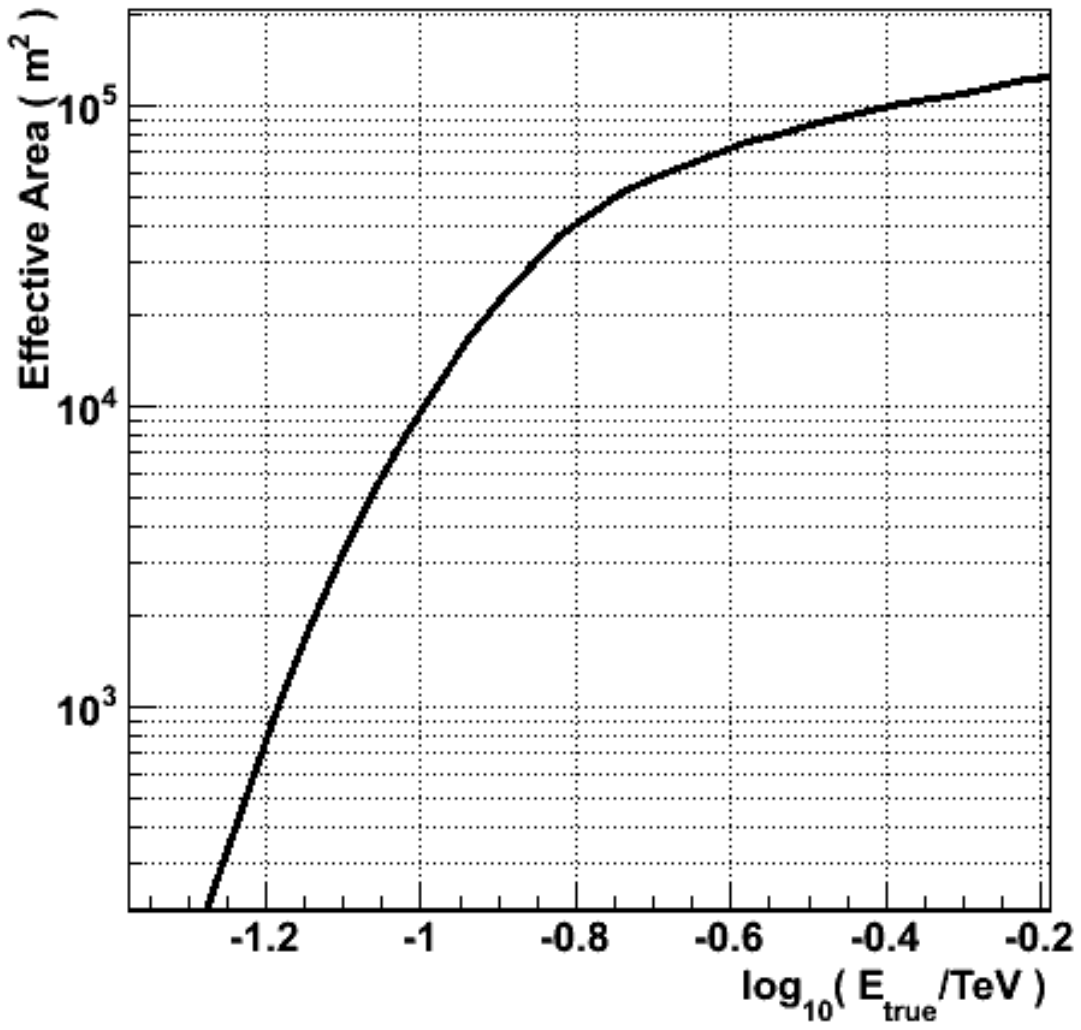

**D**

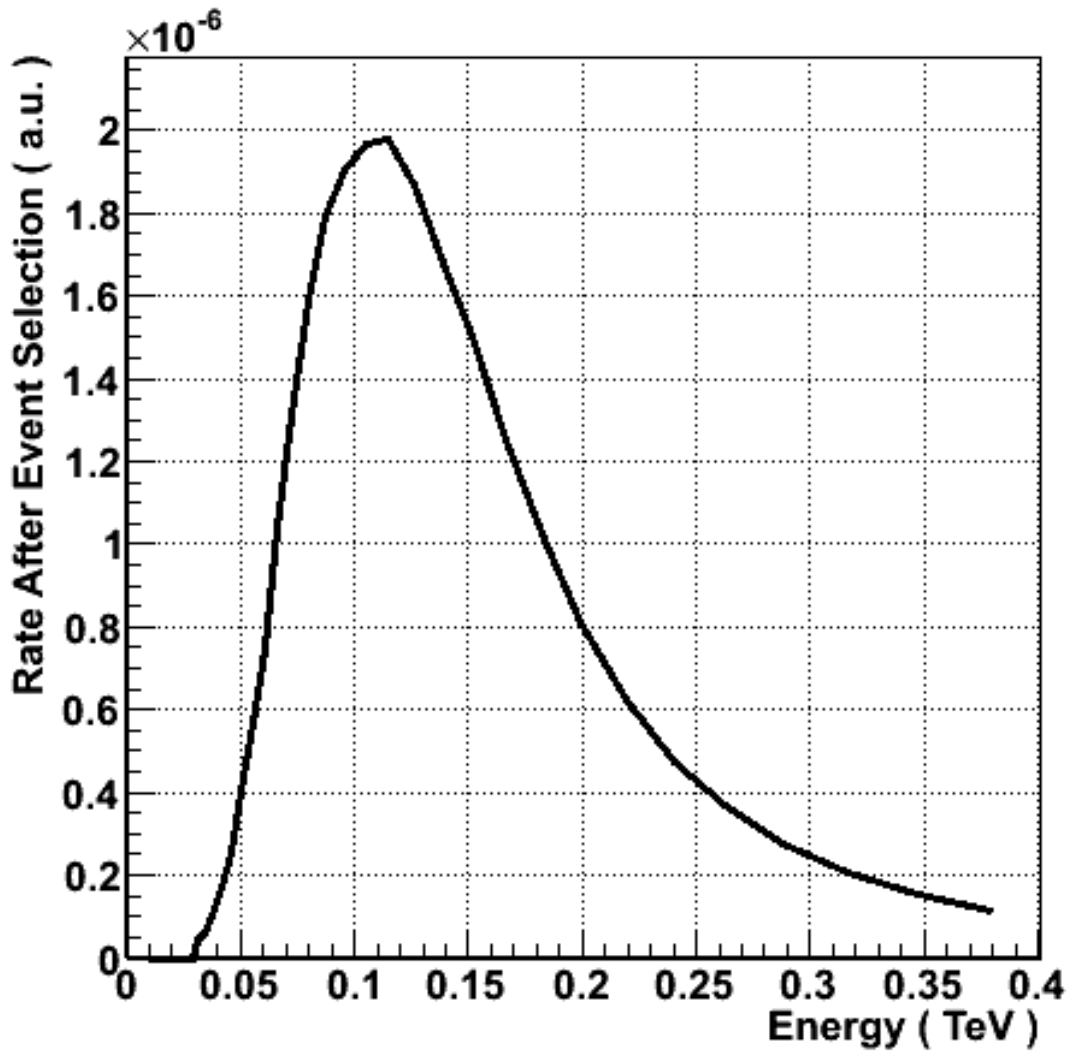

Results of the forward-folded fit of the VERITAS data with a simple power law. **A** Distribution of pulsed excess number of events versus reconstructed energy; red: VERITAS DATA, blue: best-fit distribution. The vertical solid lines mark the range of bins included in the fit. **B** $\chi^2$ dependence on the spectral index and flux normalization at 150 GeV. The cross marks the best-fit position and the solid line shows the contour line where $\chi^2$ is higher by one than the $\chi^2$ at the best-fit position (one standard deviation). **C** VERITAS effective area as a function of true energy after selection criteria. **D** Differential event rate of the best-fit Crab pulsar spectrum as a function of true energy.